\documentclass{aa}  

\usepackage{graphicx}
\usepackage{xcolor}
\usepackage{txfonts}
\usepackage{amsmath}
\usepackage{booktabs,caption}
\usepackage[flushleft]{threeparttable}
\usepackage{balance}

\def\nh{\hbox{$N_{\rm H}$}}
\def\xifu{\hbox{X-IFU}}
\def\cstat{\hbox{C-stat}}
\def\athena{\hbox{\textit{Athena}}}
\def\sixte{\hbox{SIXTE}}

\usepackage[colorlinks, citecolor=blue, urlcolor = red]{hyperref}
\makeatletter
\renewcommand*\aa@pageof{, page \thepage{} of \pageref*{LastPage}}
\makeatother

\begin{document} 

   \title{The defocused observations of bright sources with \textsl{Athena}/X-IFU}
    \titlerunning{X-IFU bright sources observations}
    
\author{E. S. Kammoun \inst{\ref{inst1},\ref{inst1a}} 
\and D. Barret \inst{\ref{inst1}} 
\and P. Peille \inst{\ref{inst2}}
\and R. Willingale \inst{\ref{inst3}}
\and T. Dauser \inst{\ref{inst4}}
\and J. Wilms \inst{\ref{inst4}}
\and M. Guainazzi \inst{\ref{inst5}}
\and J. M. Miller \inst{\ref{inst6}}
}

\institute{
IRAP, Universit\'e de Toulouse, CNRS, UPS, CNES 9, Avenue du Colonel Roche, BP 44346, F-31028, Toulouse Cedex 4, France, \email{\href{mailto:ekammoun@irap.omp.eu}{ekammoun@irap.omp.eu}}\label{inst1} 
\and
INAF -- Osservatorio Astrofisico di Arcetri, Largo Enrico Fermi 5, I-50125 Firenze, Italy\label{inst1a}
\and
Centre National d'\'{E}tudes Spatiales (CNES), 18 Avenue Edouard Belin, 31400 Toulouse Cedex 4, France \label{inst2} 
\and
Department of Physics and Astronomy, University of Leicester, Leicester, UK \label{inst3} 
\and
Dr. Karl Remeis-Observatory and Erlangen Centre for Astroparticle Physics, Sternwartstr. 7, D-96049 Bamberg, Germany \label{inst4}
\and
ESA European Space Research and Technology Centre (ESTEC), Keplerlaan 1, 2201 AZ, Noordwĳk, The Netherlands\label{inst5}
\and
Department of Astronomy, University of Michigan, 1085 South University Avenue, Ann Arbor, MI 48109, USA \label{inst6}
}


\date{Received ; accepted }

 
  \abstract{The X-ray Integral Field Unit (\xifu) is the high resolution X-ray spectrometer of ESA's \athena\ X-ray observatory. It will deliver X-ray data in the $0.2-12$~keV band with an unprecedented spectral resolution of 2.5~eV up to 7~keV. During the observation of very bright X-ray sources, the \xifu\ detectors will receive high photon rates. The count rate capability of the \xifu\ will be improved by using the defocusing option, which will enable the observations of extremely bright sources with fluxes up to $\simeq 1$~Crab. In the defocused mode, the point spread function (PSF) of the telescope will be spread over a large number of pixels. In this case, each pixel receives a small fraction of the overall flux. Due to the energy dependence of the PSF, this mode will generate energy dependent artefacts increasing with count rate if not analysed properly. To account for the degradation of the energy resolution with pulse separation in a pixel, a grading scheme (here four grades) will be defined to affect the proper energy response to each event. This will create selection effects preventing the use of the nominal Auxiliary Response File (ARF) for all events.}
  {We present a new method for the reconstruction of the  spectra obtained from observations performed with a PSF that varies as a function of energy. We apply our method to the case of the \xifu\ spectra obtained during the defocused observations.}
  {We use the end-to-end SIXTE simulator to model defocused \xifu\ observations. Then we estimate new ARF for each of the grades by calculating the effective area at the level of each pixel.}{Our method allows us to successfully reconstruct the spectra of bright sources when employed in the defocused mode, without any bias. Finally, we address how various sources of uncertainty related to our knowledge of the PSF as a function of energy affect our results.}{}


   \maketitle
%

\section{Introduction}
\label{sec:intro}

The X-ray Integral Field Unit \citep[\xifu;][]{Barret16,Barret18} is the high resolution X-ray spectrometer that is planned for launch on-board the European Space Agency \athena\ X-ray observatory \citep{Nandra13}. The instrument will deliver X-ray spectra in the $0.2-12$~keV range with a spectral resolution of 2.5~eV up to 7~keV. The overall field of view of 5\arcmin\ equivalent diameter will be subdivided into several thousand of ${\sim} 5$\arcsec\ pixels\footnote{The exact size of the pixels has not yet been decided but this will not impact the results reported in this work.}.  The main scientific goals of the instruments are 1) to study the dynamical, physical, and chemical properties of hot plasmas, notably those found in galaxy clusters, and 2) to study the extreme environment around galactic stellar-mass and supermassive black holes, accretion discs, jets, outflows and winds. In addition, the unprecedented capabilities of the \xifu\ will enable the study of many \athena\ observatory science targets, such as planets, stars, supernov\ae, compact objects, and interstellar medium. The throughput of the X-IFU decreases at high count rate. In fact, the count rate capability of the X-IFU is limited by the pixel speed, the record length required to achieve the spectral resolution, and the crosstalk level \citep[see][]{Peille18}. This capability will be improved by using the defocusing option offered by the Movable Mirror Assembly (MMA). A defocusing of 35~mm (with respect to the nominal focal length of 12~m) will enable the observations of extremely bright galactic sources with fluxes up to ${\sim} 1$~Crab \citep{Peille18}, with only limited spectral resolution degradation ($\lesssim 10$~eV). In this case, the point spread function (PSF) of the telescope will be spread over a large number of pixels so that each pixel receives a small fraction of the overall flux. Due to the energy dependence of the PSF shape, this mode will generate energy dependent artefacts increasing with count rate if not analysed properly. \cite{Peille18} focused on the ability to obtain a good enough throughput in the defocused PSF mode, without addressing the ability to reconstruct broadband energy spectra. Here, we continue this work. We present a new method that allows the analysis of \xifu\ observations of bright sources, in the defocused configuration. Whereas we focus here on the particular case of the X-IFU, we emphasize that the analysis presented below applies to any observatory with an energy dependent PSF and count rate dependent selection effects at the level of individual pixels (grading, dead time, etc...). 

\begin{figure*}
\centering
\includegraphics[width=0.6\linewidth]{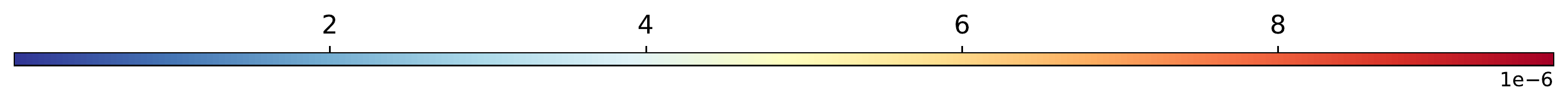}
\includegraphics[width=1\linewidth]{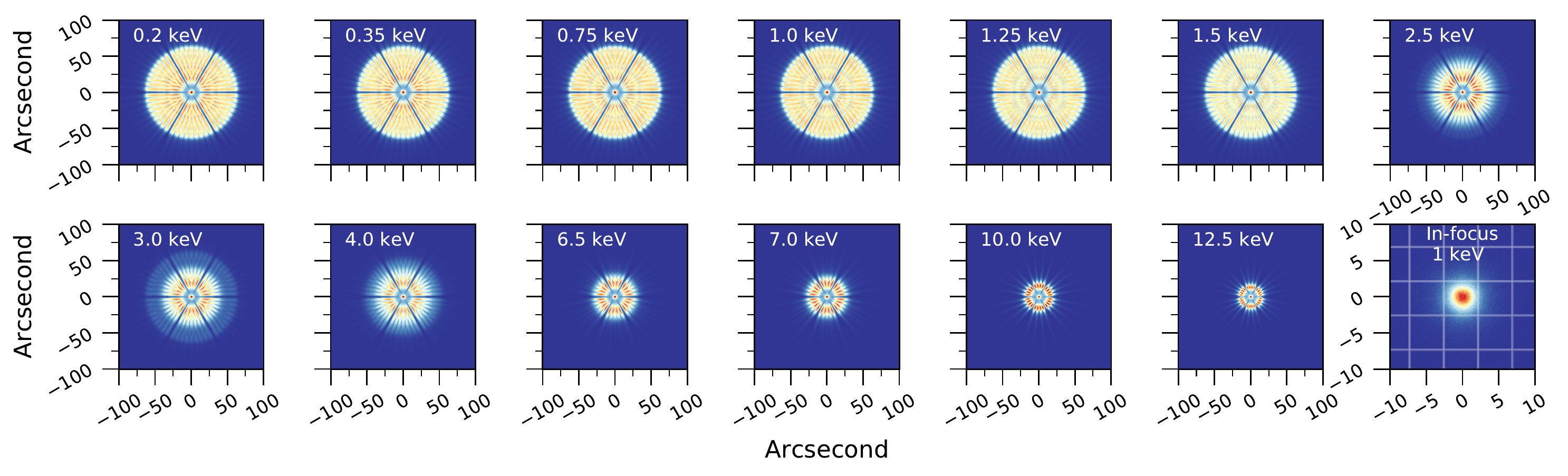}
\caption{The defocused PSF of X-IFU at 13 energies. The bottom rightmost panel shows the in-focus PSF at 1~keV ($\rm HEW = 5$\arcsec), for comparison. The grid in this pannel shows the \xifu\ pixels. We note that the in-focus PSF is shown on a $20 \arcsec \times 20\arcsec$ image while the defocused PSF is shown on a $200 \arcsec \times 200\arcsec$ image. The full field of view of the \xifu\ is of the order of 300\arcsec\ (in diameter) as shown in Fig.~\ref{fig:xifu_1kev}. All the PSFs are shown in the linear scale.}
\label{fig:defocused_psf}
\end{figure*}

In Sect.~\ref{sec:defocused}, we present the \athena\ defocused PSF. In Sect.~\ref{sec:method}, we introduce the method we developed to analyse the simulated spectra. In Sect.~\ref{sec:uncertainty}, we study the various sources of uncertainties that may affect the performance of the instrument. In Sect.~\ref{sec:groj1655}, we show an example of the ability of \xifu\ to constrain the wind absorption in bright X-ray binaries thanks to its defocusing mode. Finally, we summarise our results in Sect.~\ref{sec:discussion}.
 
\section{Defocused PSF}
\label{sec:defocused}

\begin{figure}
\centering
\includegraphics[width=1\linewidth]{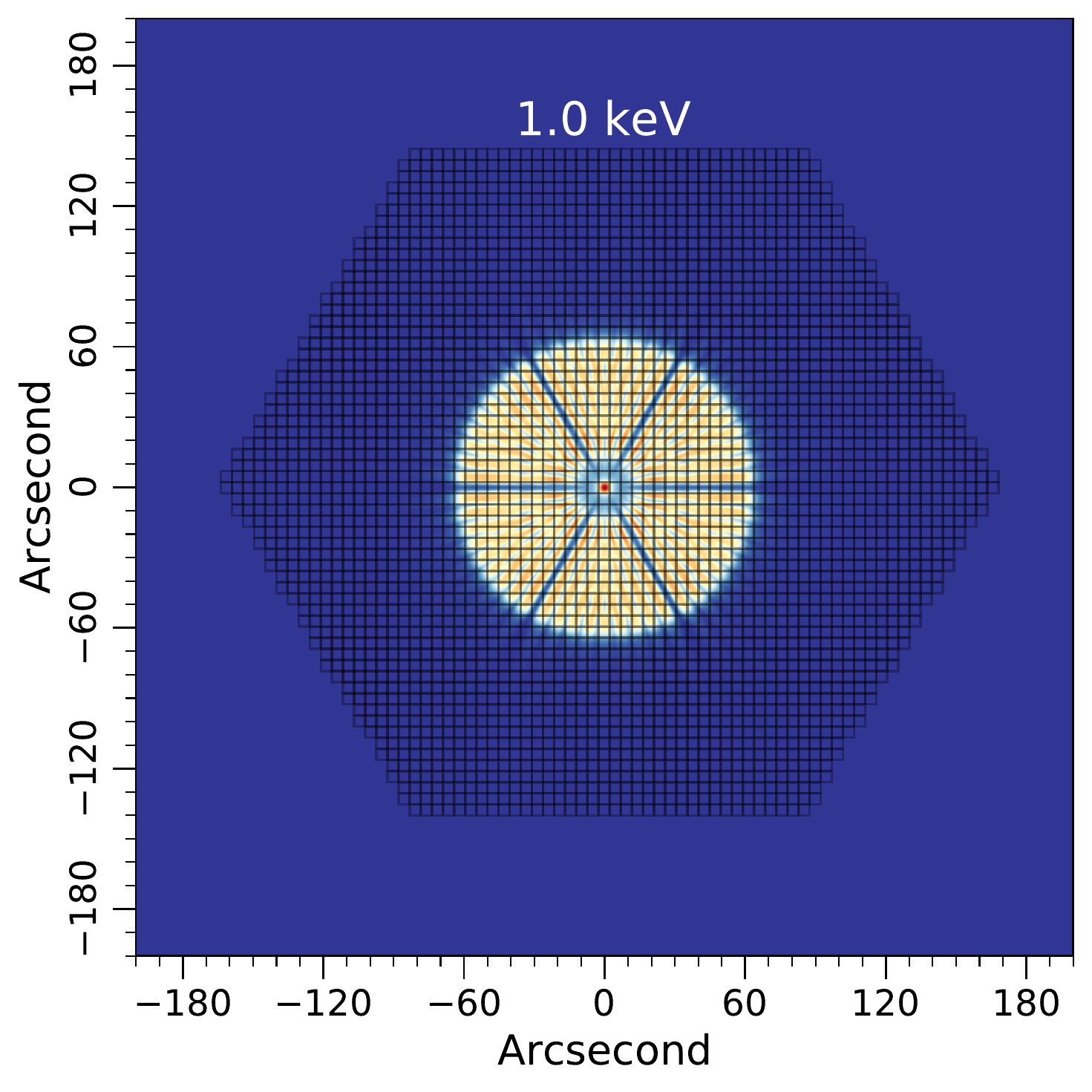} 
\caption{A zoom-out on the PSF at 1~keV, showing the full \xifu\ field of view. We note that some detailed features and the spatial variations in the PSF are happening at a level that is comparable to the \xifu\ pixel size. The same color bar as in Fig.~\ref{fig:defocused_psf} applies here.}
\label{fig:xifu_1kev}
\end{figure}

Figure~\ref{fig:defocused_psf} shows the model defocused \athena\ PSF at different energies\footnote{The PSF can be obtained from the following link \url{https://www.cosmos.esa.int/web/athena/resources-by-esa}}. The PSF is extended up to $\simeq 140$\arcsec\ in diameter in the soft X-rays. For comparison, we show the in-focus PSF at 1~keV assuming a half-energy width (HEW) of 5\arcsec, comparable to the \xifu\ pixel size. In Fig.~\ref{fig:xifu_1kev}, we show the defocused PSF at 1~keV together with the \xifu\ field of view.

The model defocused PSF is generated using a detailed physical model of the complete \athena\ mirror system. The model includes the full X-ray energy dependence of the PSF of each of the ${\sim}600$~mirror modules that populate the full aperture. The position of each module within the aperture is set such that the individual PSFs align in the focal plane to produce the in-focus PSF with HEW 5\arcsec. The telescope is defocused by shifting the mirror along the optical axis by 35~mm away from the in-focus position. In this configuration, the defocused PSF comprises a mapping of the aperture layout of the modules, including the module support structure, convolved with the PSF of the individual modules. The defocused PSF has a very strong energy dependence because the effective area of the modules depends on the radial position of each module within the aperture. Low energy X-rays ($<1.5$\,keV) are focused by all the modules but as the energy increases, the outer modules have reduced reflectivity/efficiency and at $>10$\,keV only the innermost modules produce any significant effective area.

\section{Method}
\label{sec:method}

\subsection{Simulations}
\label{sec:simulations}

We use the SImulation of X-ray TElescopes\footnote{\url{http://www.sternwarte.uni-erlangen.de/research/sixte/}} (\sixte) software package \citep{sixte} to simulate \xifu\ observations. \sixte\ is designed to perform end-to-end simulations of various X-ray observatories, including \athena. Photons are generated in a Monte-Carlo framework and then followed through the full imaging and detection process. The package uses a set of calibration files allowing to properly distribute the photon impacts, with relevant energy and timing properties, onto the focal plane. In particular, it allows us to simulate energy dependent mirror defocusing of \athena, which is the subject of this work. We note that \sixte\ adopts a linear interpolation scheme between the different energies at which the PSF is defined for lack of a more physically motivated interpolation scheme. At the end of the process, the readout energy of each event is computed through a set of response matrices and coupling mechanisms, taking into account the degradation of the instrument performance with count rate \citep[e.g., the crosstalk and the event grading; see][]{Peille18}. These are determined using \sixte's tool {\tt xifusim} \citep{Kirsch20, Lorenz20} which provides a representative simulation of the full detection pipeline of the \xifu\ including all relevant detector physics and the behavior of the readout chain.

The \xifu\ pixels are microcalorimeters: when an X-ray photon impacts the pixel, it will thermalize in an absorber thermally linked to a Transition Edge Sensor \citep[TES;][]{Smith16} microcalorimeter (originally operating at a temperature of ${\sim} 90$~mK) whose temperature will increase, leading to a rapid increase in resistance which produces a corresponding rapid decrease in current passing through the TES. This signal will be then used to estimate the energy of the event, through the application of an optimal filter \citep[e.g.,][]{Szymkowiak93, Peille16}. At high count rates, the current pulses will get packed together in the pixels’ timelines. Thus, considering an example of a pulse triplet, if the preceding event is too close to the middle event, the pulse tail from the preceding pulse leaks signal into the middle one and biases the energy estimation \citep[see][]{Peille18}. When the succeeding event is too close to the middle event, it limits the length for reconstructing the energy of the middle event, leading to a degradation of the energy resolution. To characterize these effects, four grades - namely: high resolution (HR), medium resolution (MR), limited resolution (LimR), and low resolution (LowR) - are defined in the simulator depending on the time separation of the pulses. It is worth noting that this grading scheme is applied at the level of each \xifu\ pixel.

In this work, we focus only on the effect of the grading on the broadband energy spectrum, without addressing other degradation factors that may affect the instrument performance such as pure event pile-up\footnote{The pile-up effect in the X-IFU should at a much lower level than the grading impact studied here. In all observation scenarios, the pileup fraction remains below 1\% and the undetected fraction below 0.1\% \citep[see][]{Cobo18}. } (i.e, two photons being reconstructed as one), and crosstalk. Table~\ref{tab:grade} summarizes the assumptions of time separation that we use in this work. The definition of the event grades and the responses of SIXTE are informed by detailed simulations of the detection process \citep[see e.g.,][]{Doriese09, Peille18,Kirsch20}. These definitions represent realistic assumptions, but are not the ultimate configuration that will be adopted for full operation of the instrument. In fact, the detailed X-IFU grading rules are still being optimized and are constantly evolving. As a consequence, the values reported here constitute a snapshot at the time of the simulations, explaining why they differ slightly from the earlier ones presented by \cite{Peille18}. However, we note that any changes in the definition of the grades will not affect the results of our analysis. 

\begin{table}
\centering
\caption{A typical set of parameters used for the event grade selection in this work.}
\label{tab:grade}
\begin{tabular}{llll} 
\hline \hline
Grade & Time since  & Time until &  Energy \\ 
& previous pulse &  next pulse &  resolution\\ \hline

High resolution &  7.9~ms &  52.4~ms & 2.5~eV \\
Medium resolution  & 7.9~ms  & 3.3~ms & 3~eV \\
Limited resolution  & 7.9~ms   & 1.6~ms &7~eV \\
Low resolution &  7.9~ms &  51.2~$\rm \mu s$ & $ \sim 30$~eV \\ \hline
\end{tabular}
\end{table}

The spectral analysis in this work is performed using the X-ray spectral fitting package XSPEC v12.11.1 \citep{Arnaud96}. We use the Cash statistics \citep[\cstat;][]{Cash79} to evaluate the best fits. In this work, the simulations are performed by assuming the following definition of the Crab\footnote{We note that we do not simulate the actual Crab source, which is known to be an extended source. Instead, the Crab is used to define a standard spectral shape and flux level. All the simulations used in this work assume a point-like source.}, in the XSPEC parlance:
$$ {\tt Model = TBabs \times powerlaw,} $$
\noindent where {\tt TBabs} \citep{wilms00} represents the Galactic absorption assumed to be $\nh = 4\times 10^{21}~\rm cm^{-2}$, and the power-law photon index $\Gamma = 2.1$. In this work, we consider two flux levels at 200~mCrab and 1~Crab corresponding to a normalization of 1.9 and $9.5~\rm photon~s^{-1}~cm^{-2}~keV^{-1}$ at 1 keV, respectively. These models are equivalent to an observed 2--10~keV flux of $0.41 \times 10^{-8}$ and $2.05\times 10^{-8}~\rm erg~s^{-1}~cm^{-2}$, respectively. In the case of a 1~Crab flux level, we assume a $100~\rm \mu m$ thick beryllium filter which suppresses $\sim 96\%$ and $33\%$ of the events at ${\sim} 1.5$\,keV and 3\,keV, respectively \citep{Barret18, Peille18}. This helps maximizing the throughput in the 5--8\,keV range, where disc/wind features are expected in accreting objects. We also assume the {\tt wilm} abundance and {\tt vern} cross section \citep{Verner96}. We ran the simulations with an exposure time of 2~ks and 10~ks, for each of the flux levels, assuming that the source is constant during the given exposure. 

\subsection{Reconstruction}
\label{sec:reconstruction}

\begin{figure}
\centering
\includegraphics[width=1.0\linewidth]{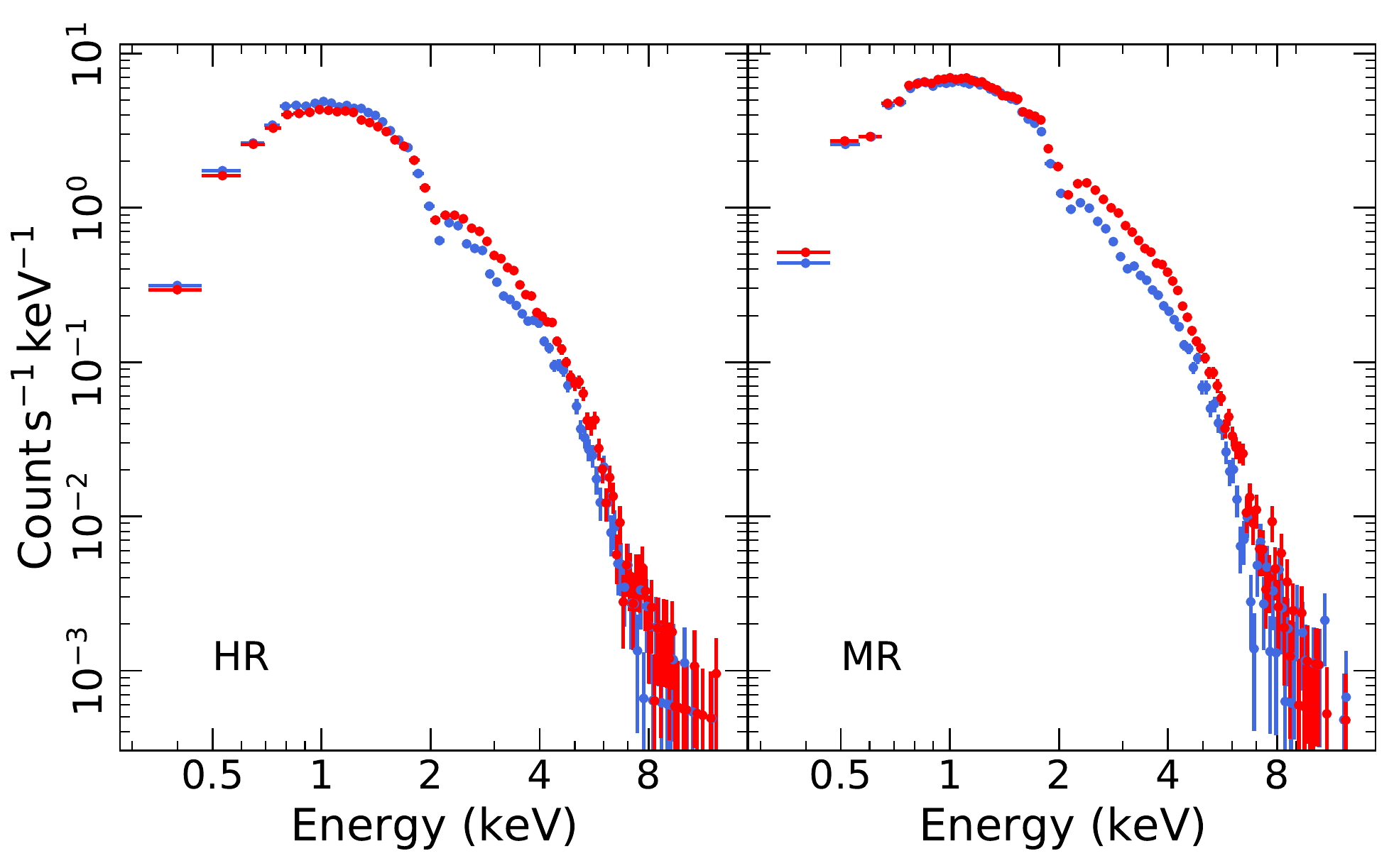}
\caption{The spectra extracted (in units of $\rm count~s^{-1}$) from two adjacent pixels (shown in red and blue) for the HR and MR event grades (left and right, respectively). The spectra are rebinned for clarity reasons.}
\label{fig:Spec_pixel}
\end{figure}

\begin{figure}
\centering
\includegraphics[width=0.9\linewidth]{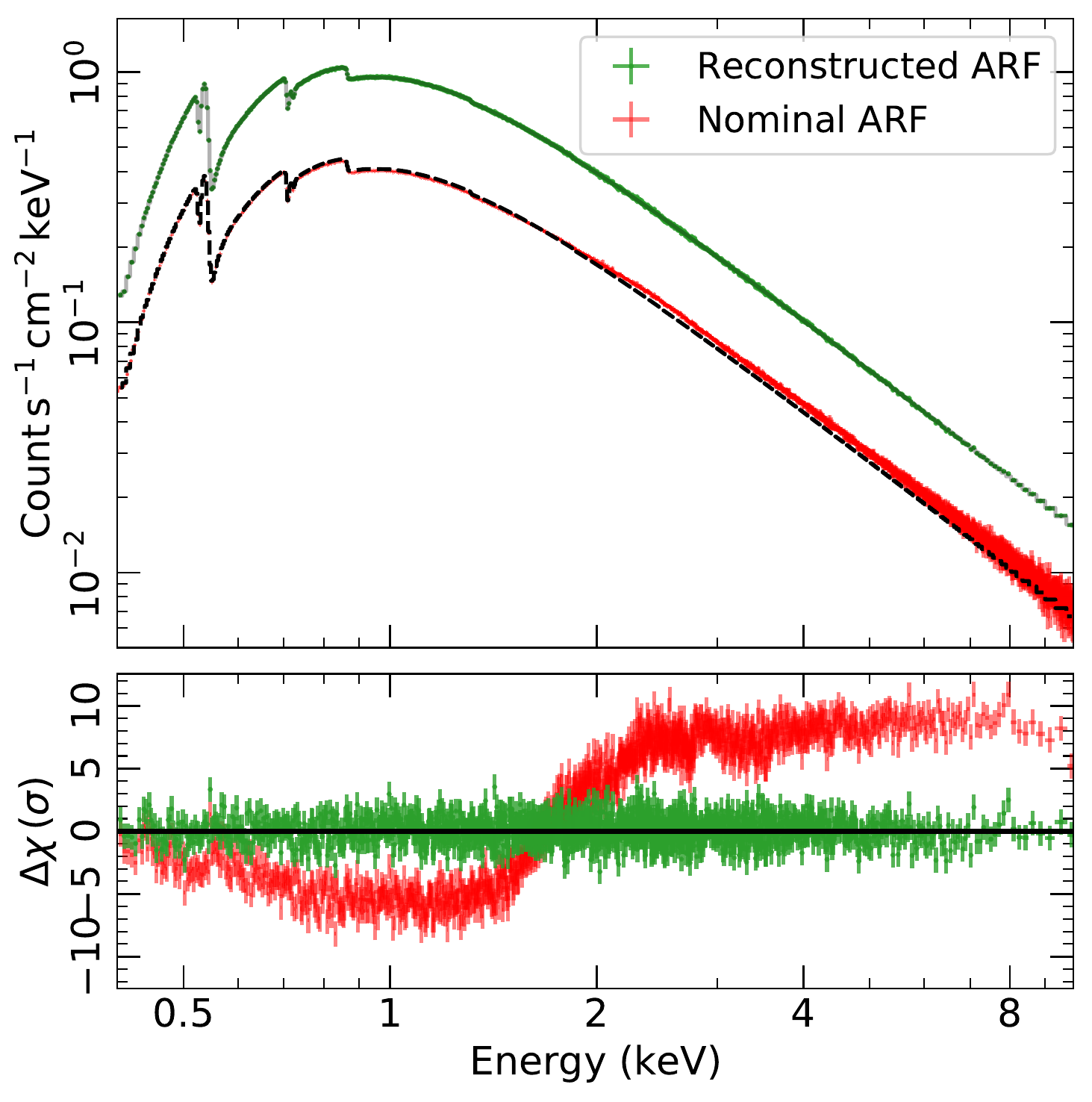}
\caption{Simulated MR spectrum assuming a flux level of 200~mCrab (with an exposure of 10\,ks), using the nominal ARF (i.e., without applying any reconstruction) in red and the reconstructed ARF in green. The black dashed line corresponds to the model obtained by fitting the spectrum using the nominal ARF, fixing \nh\ and $\Gamma$ to the input values, letting only the normalization free. We show the corresponding residuals in the bottom panel (red points). In the case of the reconstructed ARF, we show the residuals using the input parameters used in the simulations, without performing any fit. The spectra are rebinned for clarity reasons.}
\label{fig:fit_nominal_arf}
\end{figure}

\begin{figure}
\centering
\includegraphics[width=0.9\linewidth]{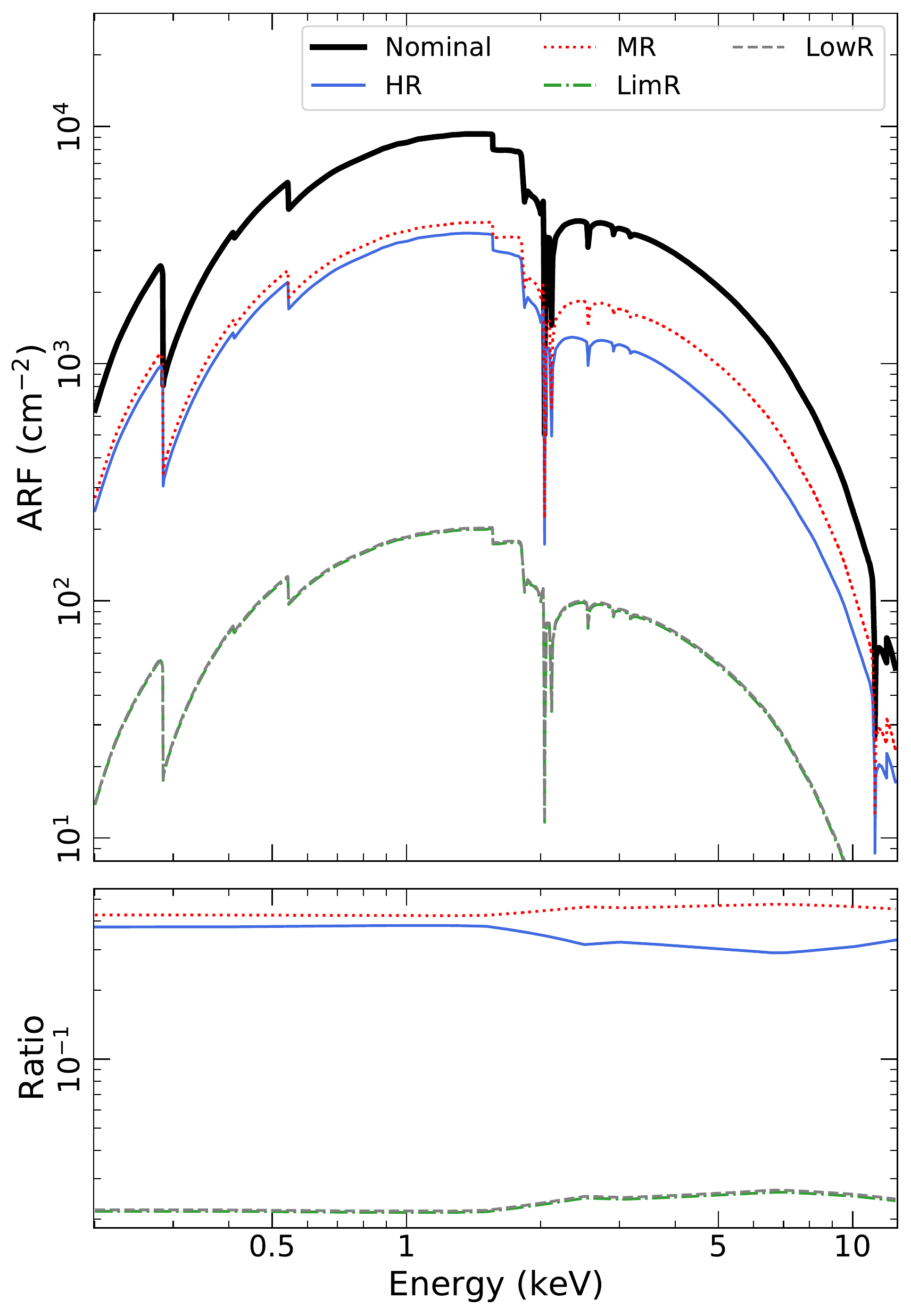}
\caption{Upper panel: The reconstructed ARFs for the various event grades. The thick solid line shows the nominal ARF for comparison. Lower panel: The ratio of the ARF at each grade divided by the nominal ARF.}
\label{fig:arfs}
\end{figure}

The output of SIXTE is a set of four spectra (one spectrum per event grade). As a result of the defocusing of the optics and the event grade scheme, the ARF of each grade will differ from the nominal ARF ($\rm ARF_{nom}$) of the instrument. The fraction of each grade in the pixels depends on the count rate the pixel detects. This depends on the shape of the PSF, and on the spectrum of the source. Thus, for each observation, a new ARF-per-grade should be calculated based on the spectrum and the flux-level of the source, which requires estimating the ARF at the pixel level. Figure~\ref{fig:Spec_pixel} shows the HR and MR spectra of two adjacent pixels, simulated assuming a flux level of 200~mCrab. This clearly demonstrates that the event grading affects each pixel differently, as it receives a different count rate. The differences between the spectra of the two pixels clearly depend on the energy, and cannot be connected by a simple normalization change. 

The red spectrum in Fig.~\ref{fig:fit_nominal_arf} corresponds to the total (i.e., the final output using all the pixels) MR spectrum obtained from the \sixte\ simulation at 200~mCrab, assuming the nominal ARF. We fitted this spectrum by fixing $N_{\rm H}$ and $\Gamma$ to the input values, and letting the normalization free to vary. The fit is not statistically accepted ($\mathrm{C-stat/dof=4.4}$), and results in a normalization that is $\sim 2.4$ times smaller than the input value. The bottom panel of Fig.~\ref{fig:fit_nominal_arf} shows clear energy-dependent residuals that result from the use of the nominal ARF. Similar erroneous fits and strong residuals can be seen for all of the event grades.

In the following, we present a new, generic technique that allows to estimate the ARF for each grade, taking into consideration all the aforementioned effects. We stress that this method can be applied to the observations of any instrument operated out of focus and/or whose PSF depends on the energy. This method is thus independent of shape of the PSF. In this work, we apply this method to the particular case of the \xifu. The simulations performed using \sixte\ are intended to replicate a real observation by the \xifu. We note that in the absence of a physical, parametric, model of the PSF, we assume in the following a perfect knowledge of the PSF based on the description provided in Sect.~\ref{sec:defocused}. Such physical model of the PSF will be based on measurements obtained from the calibration of the \athena\ mirror. Here are the steps we follow to reconstruct the spectrum at each of the grades:

\begin{figure*}
\centering
\includegraphics[width=0.45\linewidth]{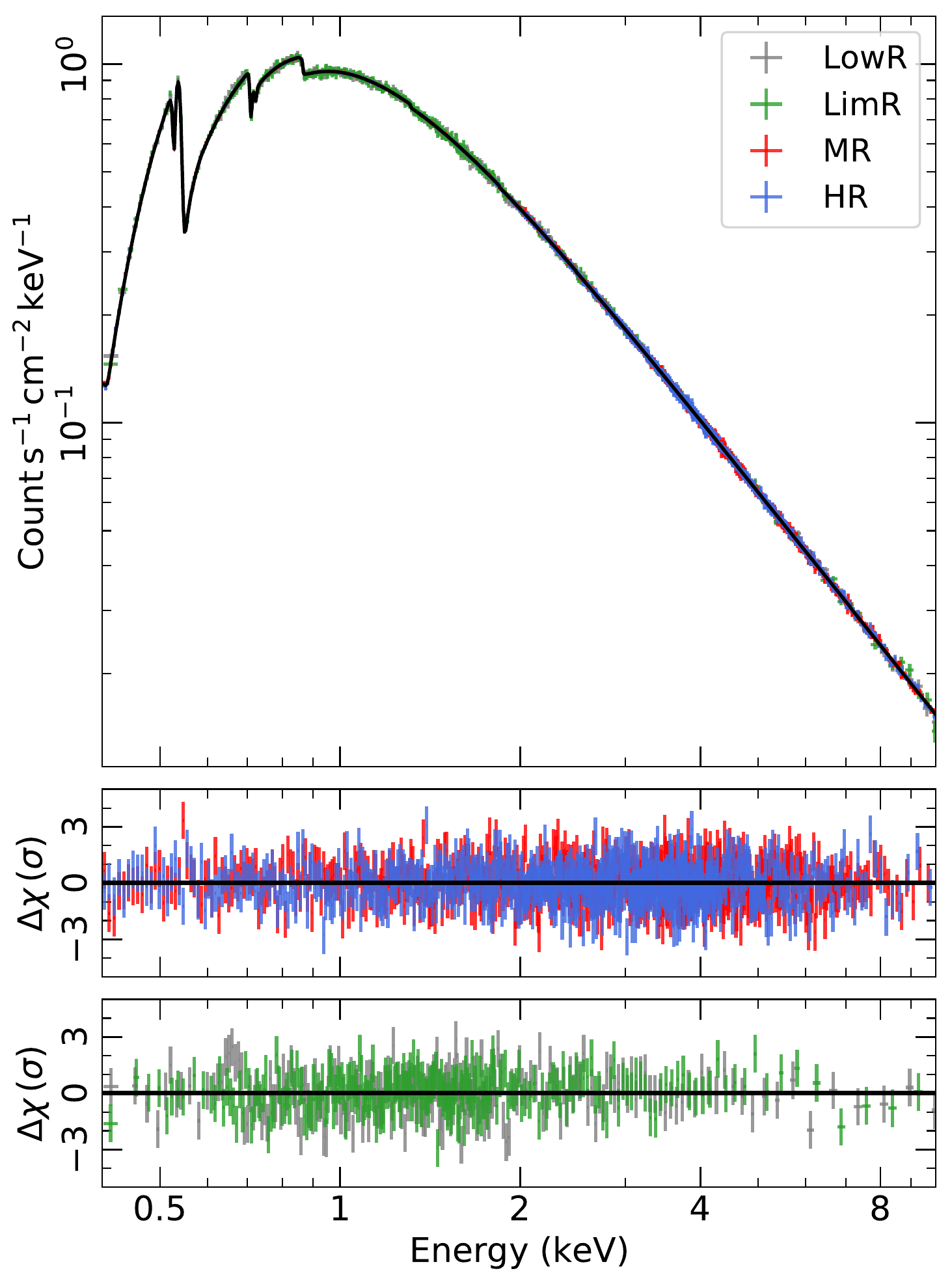}
\includegraphics[width=0.45\linewidth]{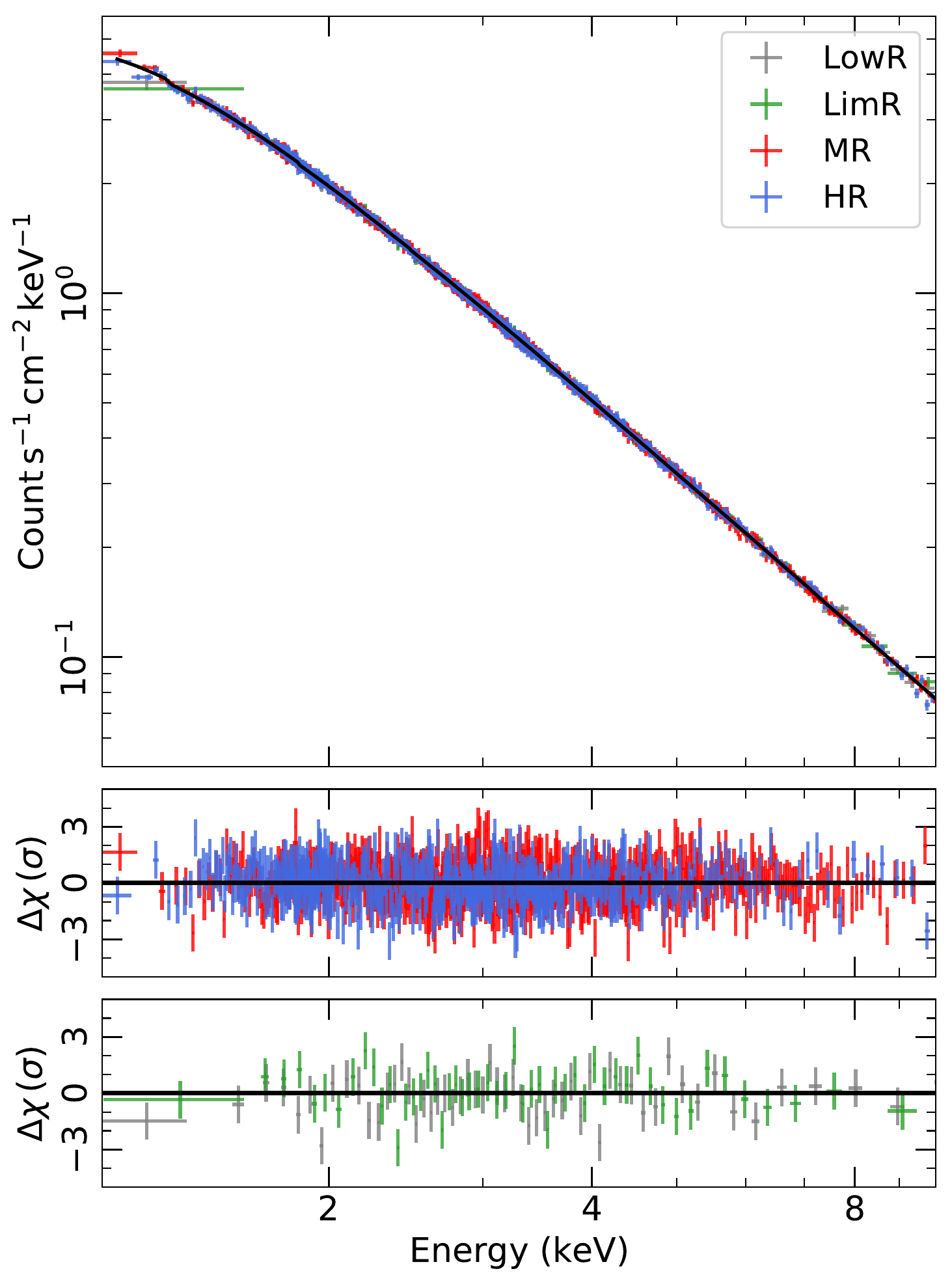}
\caption{The reconstructed simulated spectra (normalised by the effective area) assuming flux levels of 200~mCrab (left) and 1~Crab (right) for the different event grades (HR: blue, MR: red, LimR: green, LowR: grey), assuming an exposure of 10\,ks. In the latter case, we use a $100~\rm \mu m$ Be filter. The bottom panels show the residuals for each event grade. All the spectra are rebinned for clarity reasons.}
\label{fig:200mcrab_1crab}
\end{figure*}

\begin{itemize}
\item[] {\bf 1. Estimating the PSF and ARF at each of the pixels.} As mentioned before, the grading is applied at the pixel level. Thus, we need to estimate the PSF at a pixel ($i$), at all energies, $\rm PSF_{i} (E)$. To do so, we integrate the PSF within each of the \xifu\ pixel at each of the energies where the PSF is defined. 

As mentioned earlier, we assume a perfect knowledge of the instrument PSF. Thus, in order to estimate $\mathrm{PSF_i}$ in the energy ranges where the model PSF is not actually defined (see Sect.~\ref{sec:defocused}), we \textit{linearly interpolate} over the full energy range to replicate the particular way in which \sixte\ runs the simulations.

Then, we estimate the ARF at each of the pixels by multiplying the nominal ARF by the PSF at each pixel: 
$$\rm ARF_{i} (E) = PSF_{i}(E) \times ARF_{nom}(E).$$\\

\item[] {\bf 2. Estimating the total number of counts per pixel ($\mathrm{CR}_{\rm tot,i}$).} The event grading applied at each pixel depends on the total count rate that is detected in each pixel. In the case of SIXTE, the simulator provides the number of valid events ($\mathrm{CR}_{\rm val, i}$), which corresponds to the number of generated photons that are within the energy band $0.2 - 12$~keV. To estimate $\mathrm{CR}_{\rm tot,i}$, we numerically solved the following equation:

$$\mathrm{CR}_{\rm val, i} = \mathrm{CR}_{\rm tot,i} \exp \left(-\mathrm{CR}_{\rm tot,i} \times \delta t \right), $$
\noindent where $\delta t$ is the minimum time separation with respect to a previous pulse for an event to be valid as defined in Table~\ref{tab:grade} ($\delta t= 7.9~$ms, which corresponds to 10 times the fall time of the X-ray pulses. ).\\

\item[] {\bf 3. Estimating the fraction of events of each grade in each pixel.} As the arrival time of X-ray events follow Poisson statistics, we adopt the following top down exclusion scheme \citep[see e.g.,][]{Seta12, Peille18}:

 \begin{align}
     w_{\rm HR, i} &= e^{- \Delta T_{\rm HR} \times \mathrm{CR}_{\rm tot,i} }\\
    w_{\rm MR, i} &=  e^{- \Delta T_{\rm MR} \times \mathrm{CR}_{\rm tot,i}} - e^{- \Delta T_{\rm HR} \times \mathrm{CR}_{\rm tot,i} } \\
    w_{\rm LimR, i} &= e^{- \Delta T_{\rm LimR} \times \mathrm{CR}_{\rm tot,i} } - e^{- \Delta T_{\rm MR} \times \mathrm{CR}_{\rm tot,i} } \\
    w_{\rm LowR, i} &= e^{- \Delta T_{\rm LowR} \times \mathrm{CR}_{\rm tot,i} } - e^{- \Delta T_{\rm LimR} \times \mathrm{CR}_{\rm tot,i} }
\end{align}
 
\noindent where $\Delta T_{\rm HR}$, $\Delta T_{\rm MR}$, $\Delta T_{\rm LimR}$, and $\Delta T_{\rm LowR}$ are the sum of the grading criteria in both columns of Table~\ref{tab:grade}, for each of the event grades.\\

\item[] {\bf 4. Calculating the total ARF for each of the event grades ($\rm  ARF_{g}$),} being the sum of all $\rm ARF_i$ weighted by the fraction of events detected at each grade,
$$ {\rm ARF_{g}(E)} =  \sum_i w_{\rm g, i} \times{\rm  ARF_{i}(E) } .$$
We calculate the sum using the FTOOLS command {\tt addarf}.
\end{itemize}

\noindent The end result of this approach are four ARFs, one for each of the event grades. The different ARFs calculated for a simulation of 200~mCrab are shown in Fig.~\ref{fig:arfs}.

\begin{figure}
\centering
\includegraphics[width=0.95\linewidth]{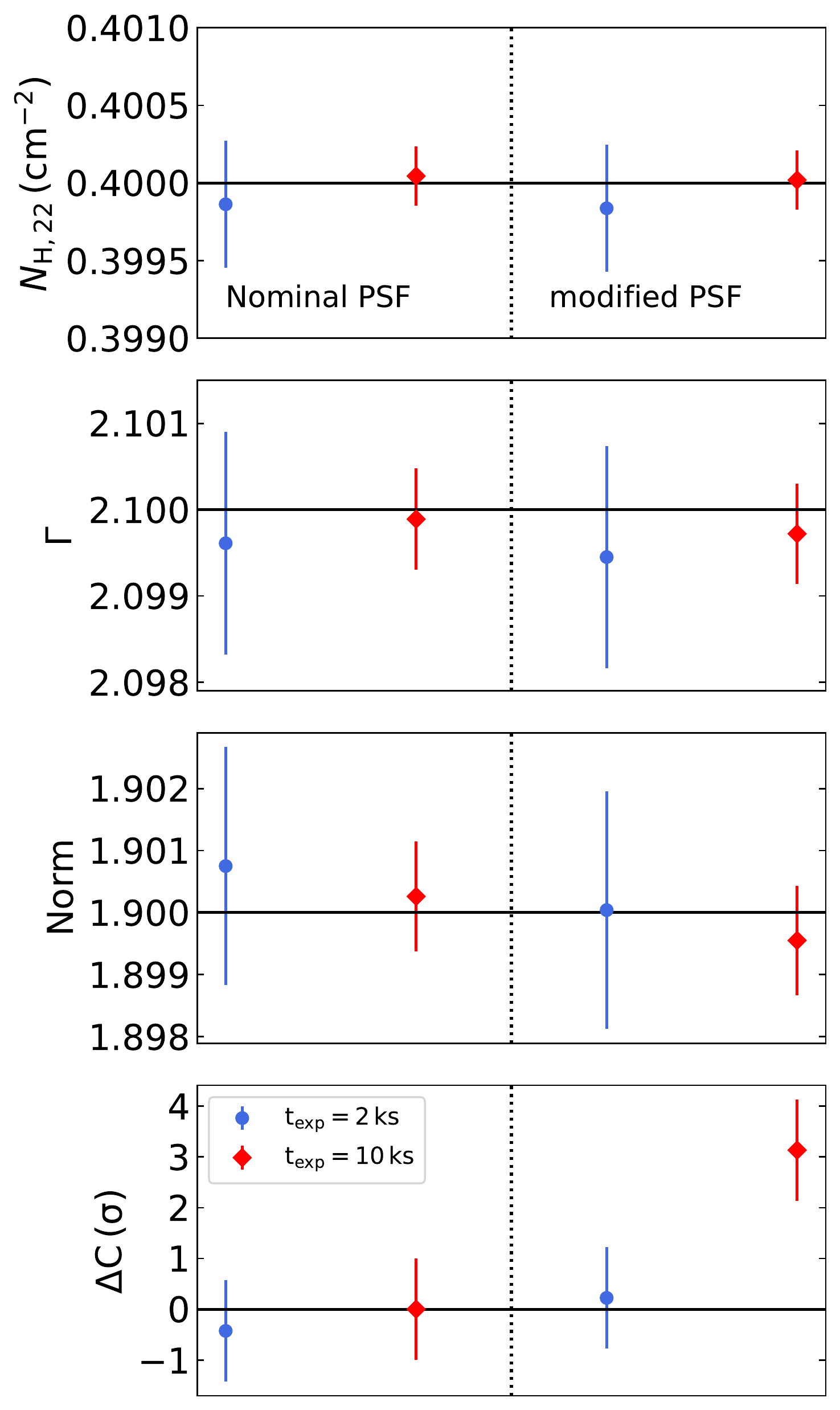}
\caption{The best-fit parameters obtained by fitting simultaneously all the event grades for a flux level of 200~mCrab using the nominal PSF (left) and the modified PSF (right, as explained is Sect.~\ref{sec:psfmodified}), for $t_{\rm exp} = 2$~ks and 10~ks (blue circles and red diamonds, respectively). The error bars correspond to the 90\% confidence level. The bottom panel shows the deviation of the best-fit \cstat\ from the expected value estimated by following the analytic prescription presented by \cite{Kaastra17}.}
\label{fig:200mcrab_bestfit}
\end{figure}

\subsection{Spectral fitting}
The green spectrum in Fig.~\ref{fig:fit_nominal_arf} represents the MR spectrum using the reconstructed ARF in green. The green residuals shown in the bottom panel of this figure correspond to the input model parameters, without fitting. Contrary to the red spectrum, where the nominal ARF is used, the residuals are flat (consistent with zero) not showing any systematic. This figure demonstrates that the differences between the two spectra cannot be explained by a simple normalization factor. Instead, strong energy-dependent features can be seen in the non-reconstructed spectrum.

In Fig.~\ref{fig:200mcrab_1crab}, we show the spectra of all the grades for the two flux levels, with an exposure time of 10~ks. We also show the input model, without any fitting. The model is statistically accepted and does not show any systematic residuals which arose when using the nominal ARF. In the bottom panels of Fig.~\ref{fig:200mcrab_1crab}, we show the residuals for each of the event grade. We note that in the case of the 200~mCrab spectrum, we increased the number of energy bins of the model to be able to account for all the features in the spectrum, using the XSPEC command {\tt energies}. The left panel of Fig.~\ref{fig:200mcrab_bestfit} shows the best-fit parameters for the 200-mCrab observations, with $t_{\rm exp} = 2$\,ks an 10\,ks. The best-fit parameters are all consistent with the input values.  In order to test the goodness of the fit, we compare the best-fit \cstat\ to the expected \cstat\ and its corresponding standard deviation, following the analytic prescription by \cite{Kaastra17}. For both cases, the observed \cstat\ is consistent with the expected one within less than 1$\sigma$.

\section{Impact of the PSF calibration uncertainties}
\label{sec:uncertainty}

The reconstruction method defined above assumes a perfect knowledge of the PSF. However, the final performance of the instrument, and the effectiveness of the reconstruction method, both heavily rely on the degree of knowledge of the PSF. In this section, we explore the different sources of uncertainties that may affect the reconstruction of the ARFs. Notably, we address the uncertainties that are related to the knowledge of the PSF and its energy dependency, within the current framework in which an analytical model of the PSF is used. The tests performed in this section use a toy model for the PSF uncertainties, in the absence of a parametric model of the PSF.

\subsection{PSF calibration as a function of energy}
\label{sec:interpolation}

\begin{figure}
\centering
\includegraphics[width=0.98\linewidth]{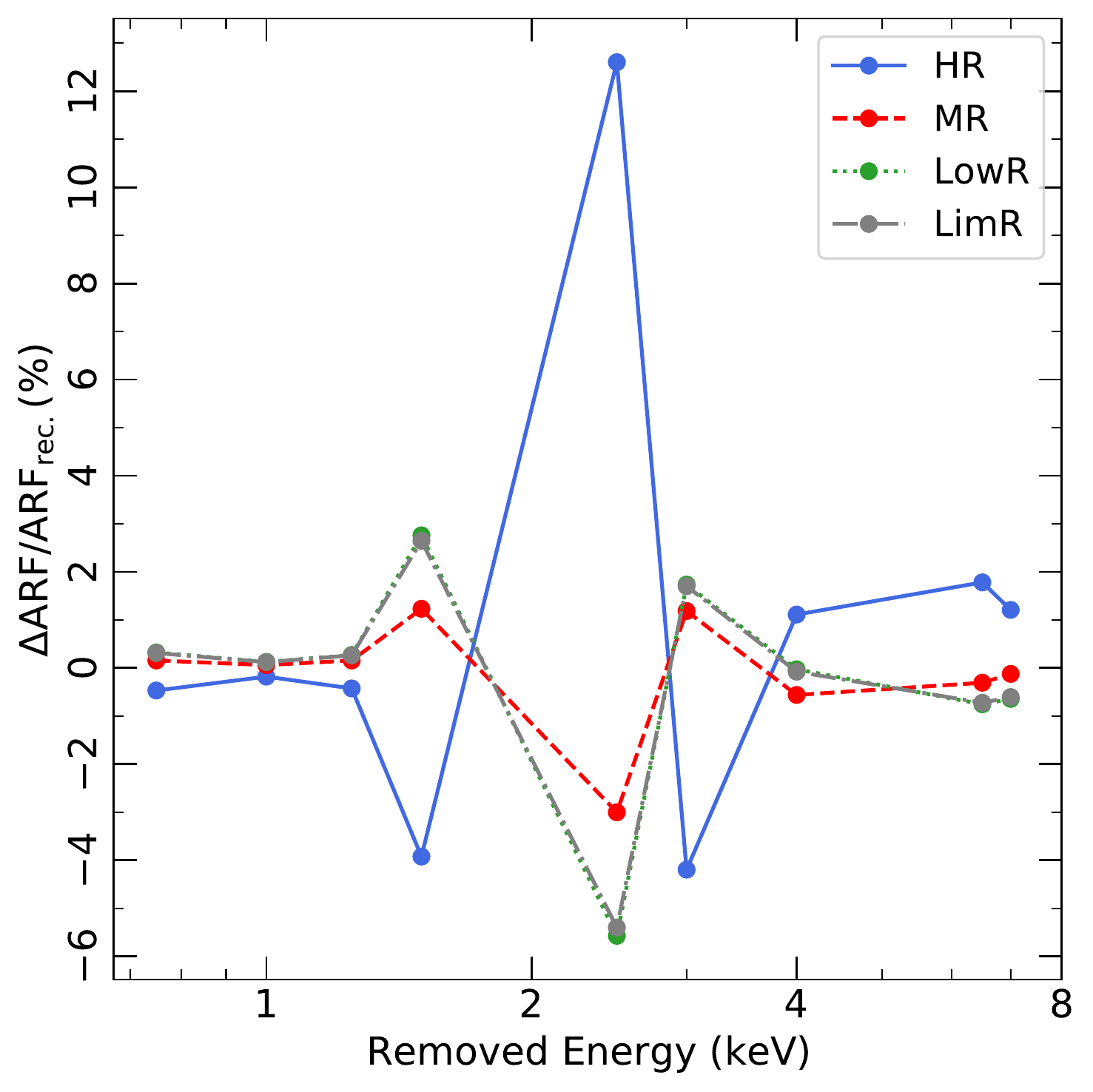}
\caption{The percent difference between the ARF obtained by removing an energy during the reconstruction and the ARF using all energies as a function of the removed energy, for the different event grades.}
\label{fig:removing_energy}
\end{figure}

\begin{figure*}
\centering
\includegraphics[width=0.9\linewidth]{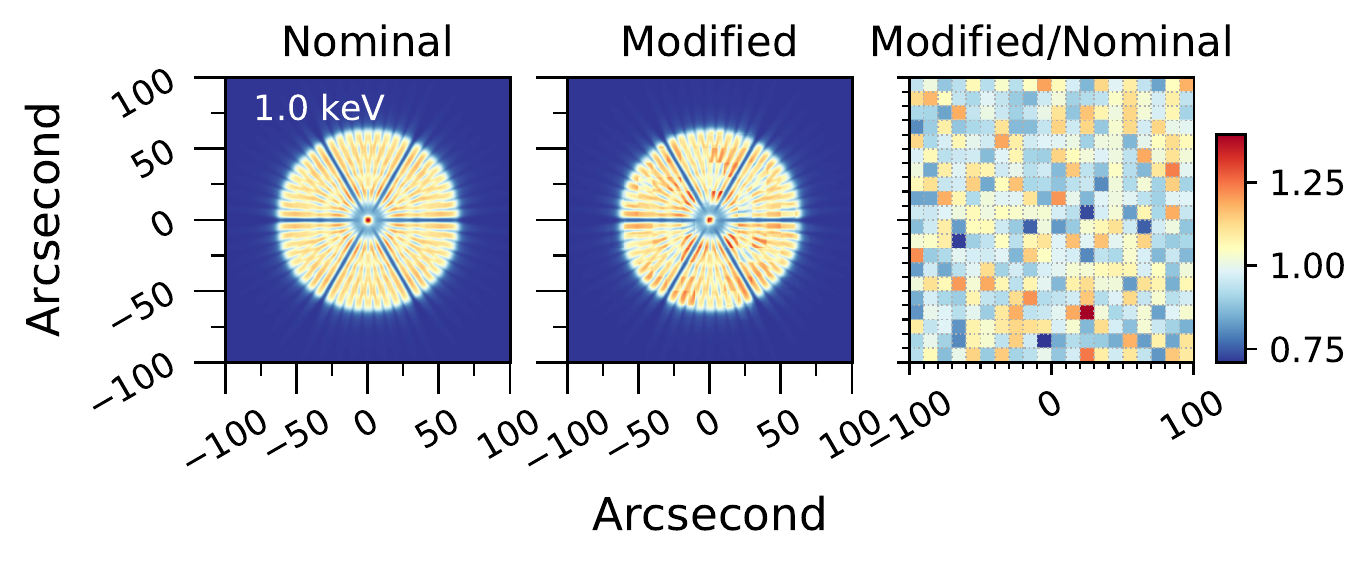}
\caption{Left: Nominal defocused PSF at 1~keV. Middle: The modified PSF at 1~keV, estimated by including an uncertainty of with a standard deviation of 10\%\ on the value of the PSF in bins of 10\arcsec. Right: The ratio of the modified to nominal PSF value. The colorbar corresponds to the ratio in the right panel.}
\label{fig:modifiedPSF_1kev}
\end{figure*}

In this section, we investigate the uncertainties that may be introduced in the ARF reconstruction due to the lack of knowledge of the PSF at a given energy. In this case, we use the simulations presented in Sect.~\ref{sec:reconstruction}, for the 200~mCrab level, that are performed assuming a knowledge of the PSF at all 13 energies. Then, during the reconstruction we assume that in Step~1, the PSF is known at 12 energies only. In other terms, the interpolation of the PSF as function of energy is performed by omitting one of the energies in the range $[0.75-7]$~keV.

Figure~\ref{fig:removing_energy} shows the percent difference in ARF that is introduced by removing one of the energies during the interpolation for the various event grades. The difference is significant when we removed energies in the ${\sim} 1.25-3$~keV range, in all event grades (reaching $\sim 12\%$ at 2.5\,keV for the HR). This is due to the fact that, in this energy range, the PSF significantly changes shape by going from one energy to another. Thus, by omitting one intermediate energy, the interpolation scheme (in Step~1) will not be able to account for the change in the PSF shape, which will result in a wrong estimate of the effective area around the omitted energy. We present in Fig.~\ref{figapp:removing_energy} the residuals obtained by removing each of those energies during the reconstruction, for all the event grades. We also show in the bottom panel of the same figure, how the \cstat\ changes by removing each of those energies. We note that the residuals and the \cstat\ values reported in this figure are estimated by comparing the spectra to the input model, without fitting. Strong systematic residuals (in all event grades) are observed in the ${\sim} 1.25-3$\,keV range, where the difference in the ARF is the largest. These residuals are not random, but appear to be a systematic excess/deficiency in a given energy range. It is also worth noting that these effects are local, and do not affect the global parameters of the continuum. In fact, the residuals are flat outside the ranges that are affected by omitting a given energy. However, these residuals could be confused with emission/absorption features in the ranges of interest.

\subsection{Uncertainty on the PSF}
\label{sec:psfmodified}

\begin{figure}
\centering
\includegraphics[width=0.95\linewidth]{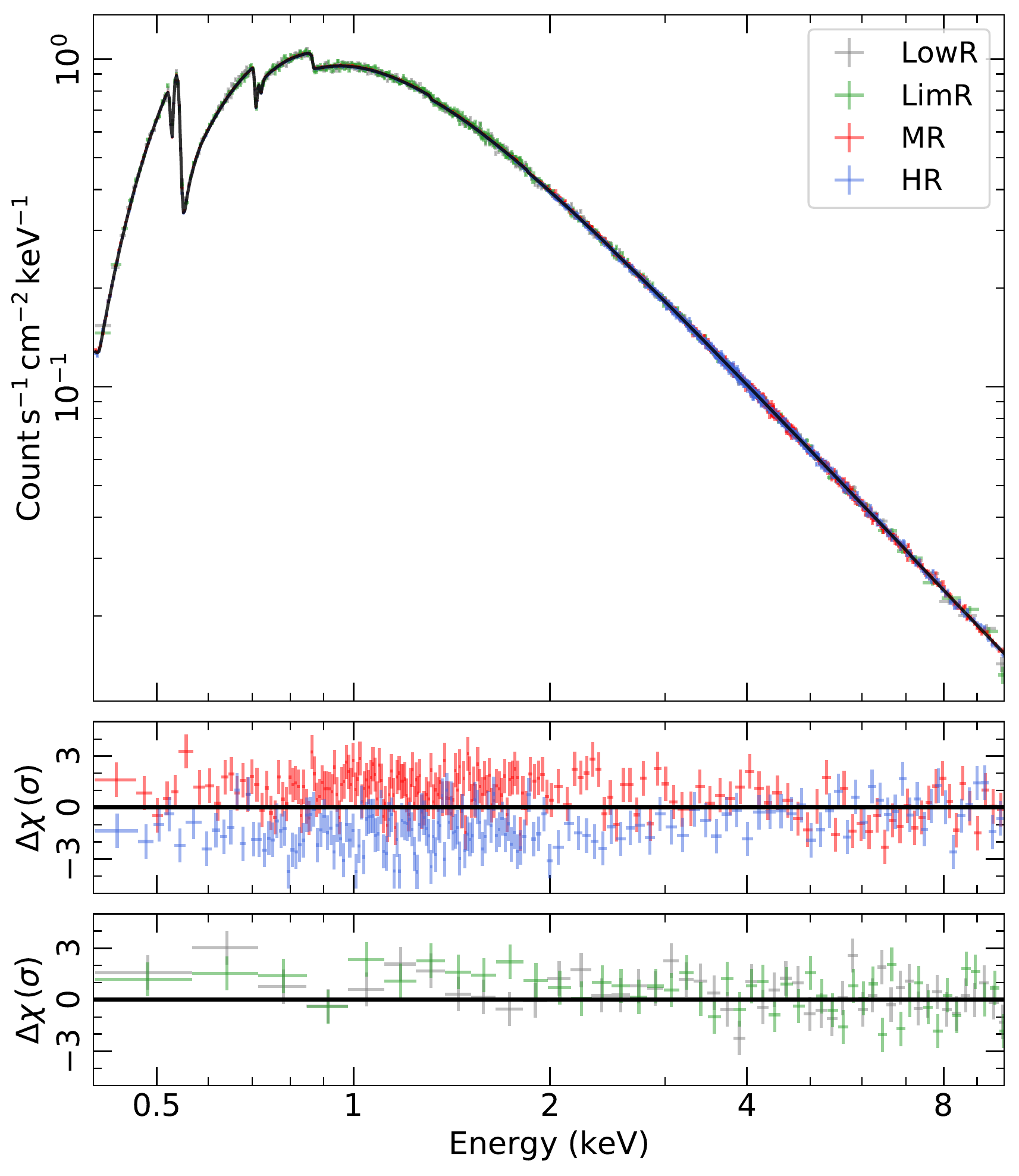}
\caption{The reconstructed simulated spectra (normalised by the effective area) assuming a flux level of 200~mCrab and an exposure of 10\,ks, for the different event grades, using the modified PSF as explained in Sect.~\ref{sec:psfmodified}. The bottom panels show the residuals for each event grade. All the spectra are rebinned for clarity reasons.}
\label{fig:200mcrab_modifiedPSF}
\end{figure}

\begin{figure}
\centering
\includegraphics[width=0.85\linewidth]{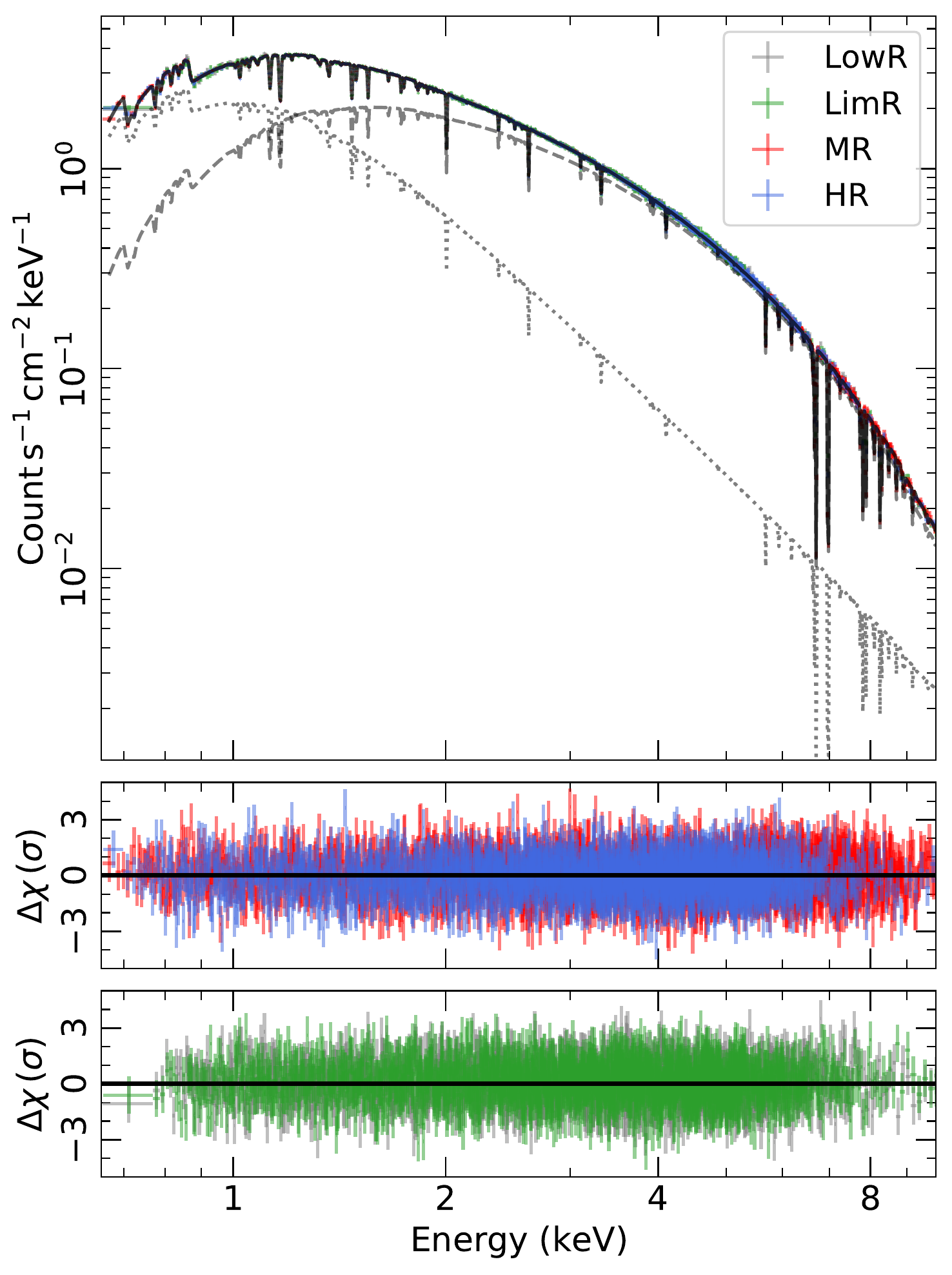}\\
\includegraphics[width=0.85\linewidth]{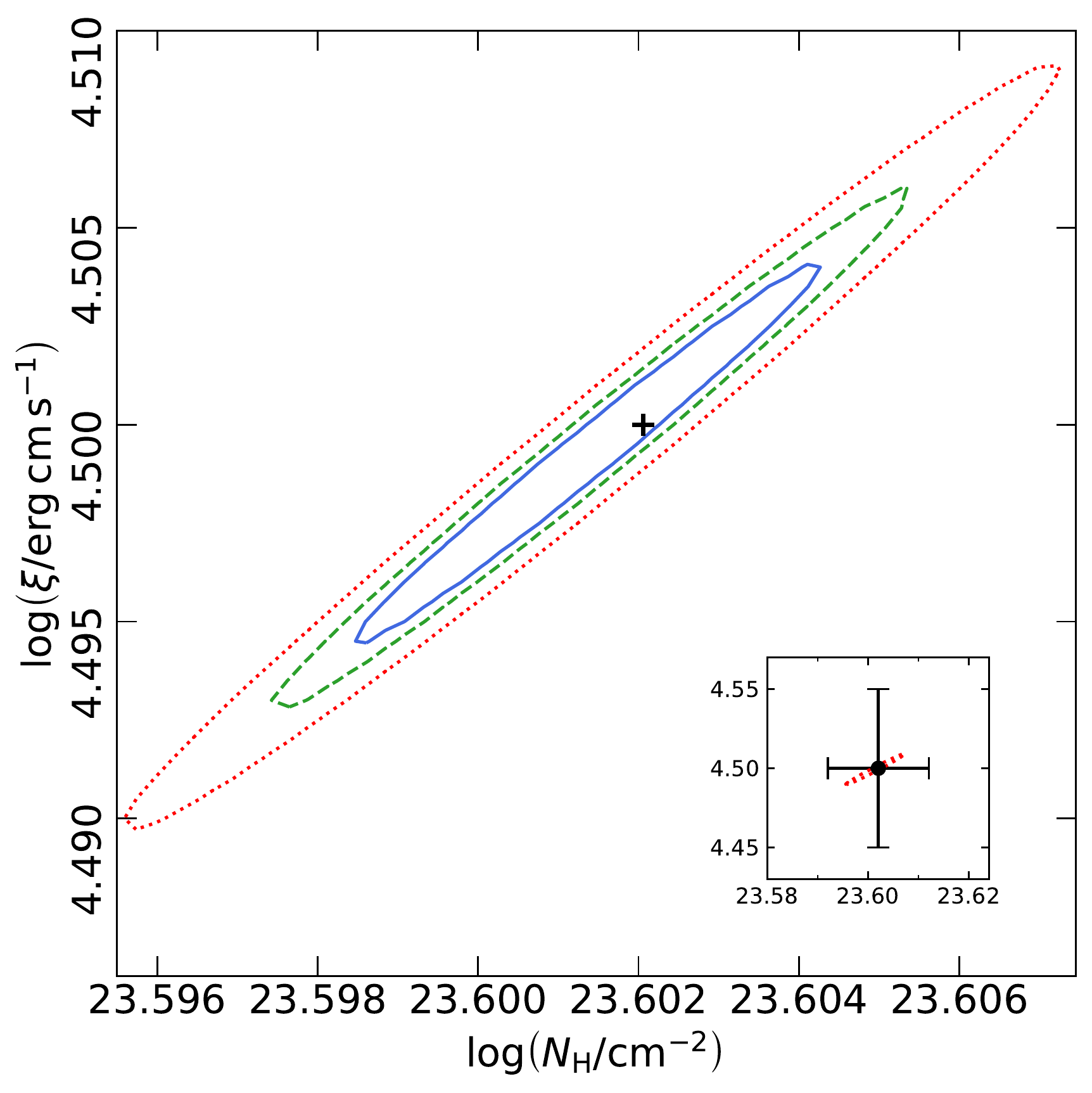}

\caption{Top: The reconstructed simulated spectra of GRO~J1655$-$40 for $t_{\rm exp} = 10$~ks at a flux level of 1~Crab. The dotted and dashed lines represent the blackbody and the power-law components, respectively. The spectra are rebinned for clarity reasons. Bottom: the $\log \xi$ vs. $\log N_{\rm H}$ confidence contours at the 1, 2, and $3\sigma$ levels. The black cross shows the input values used for the simulations. The inset shows the typical precision (at $1\sigma$) that can be achieved using current facilities. We plot the $3\sigma$ contours obtained with the X-IFU, for comparison.}
\label{fig:groj1655}
\end{figure}
We investigate in this section the effects on the reconstructed ARFs introduced by the variation in the effective areas from each PSF modules in the mirror aperture. In fact, the modules are arranged in 15 rows (annular rings) in the aperture. The model assumes that the modules within each row are all identical and have exactly the same effective area and PSF as a function of energy. The resulting model PSFs are the ones that have been used in the previous sections. However, in reality the effective area and PSF of the modules will vary because of manufacturing tolerances and imperfections in the reflecting surface coatings. Thus, it is expected that the real effective area and PSF will show variations at the level of the modules. These variations will modulate the brightness of the individual blobs in the defocused PSF. To simulate these effects, we randomize the value of the PSF at each of the module. We consider a grid of $10\arcsec \times 10\arcsec$ that is equivalent to the module size, and we randomize the value of the PSF in each of the grid by assuming a normal distribution around the nominal PSF value with a standard deviation of 10\%. The choice of the scale size is dictated by the angular resolution ($2\times \rm HEW$ of the in-focus PSF). In fact, the area cannot vary over a scale size that is less than the angular resolution. The amplitude of the variations in effective area is controlled by the error budget tolerances used in the manufacturing process. Using a standard deviation of 10\% with a normal distribution is a conservative upper limit  to the expected variations over the modules and is in line with the top level requirements on the absolute effective area of \athena\ mirror. In Fig.~\ref{fig:modifiedPSF_1kev}, we show the nominal (model) and modified defocused PSF at 1~keV. We also show the ratio of the two PSFs.

Then, we reconstructed the ARFs for all grades using the modified PSF. Figure~\ref{fig:200mcrab_modifiedPSF} shows the reconstructed spectra using the modified PSF, for the 200-mCrab simulation with $t_{\rm exp} = 10$\,ks. The residuals show a systematic trend notably below ${\sim} 3$~keV in all grades. In the right-hand side of Fig.~\ref{fig:200mcrab_bestfit}, we show the best-fit results and the deviation of the \cstat\ from the expected value, for $t_{\rm exp}=2$\,ks and 10\,ks. It is clear from this figure that using the modified PSF, the best-fit \cstat\ is more than 3$\sigma$ larger than the expected value, for $t_{\rm exp}=10$~ks, resulting in an unaccepted fit. However, the fit is statistically accepted for the lower exposure time simulation. Our results show that our capabilities to observe bright sources with the \xifu\ depend on our knowledge of the variations in effective area for each individual module, especially for the deepest exposures.

\section{Application: The case of GRO~J1655--40}
\label{sec:groj1655}

In this section, we show an example of the capabilities of \xifu\ in studying bright sources. We use GRO~J1655$-$40 as source for our simulations. It is a typical bright, hard state black hole with strong wind signatures and can be regarded as a canonical source to study absorption lines caused by accretion disk winds. We use a model, with spectral parameters taken from \citet{Miller08}, which consists of a neutral Galactic absorption in the line-of-sight of the source, in addition to a strong ionized absorption from a wind (with a column density and ionisation parameter of $N_{\rm H} = 4\times 10^{23}~\rm cm^{-2}$ and $\log \left(\xi/\rm erg~cm~s^{-1} \right) = 4.5 $, respectively) that acts on the sum of a disk blackbody (with a temperature $kT_{\rm in} = 1.35$\,keV) and a powerlaw (with a photon index $\Gamma =3.5$). We assume a source flux that is equivalent to 1~Crab with an exposure time of 10~ks. 

First, we fitted the simulated spectra using the nominal ARF (i.e., without reconstruction), simultaneously for all grades. We tied the parameters between all of the grades. The fit is not statistically acceptable (\cstat$\rm /dof=22143$), showing strong energy-dependent residuals. In addition, the best-fit parameters of the continuum components (blackbody and power law) deviate from the input parameters as shown in Table~\ref{tab:fitgro}. In particular, the normalization relative to each of those components deviates from the input value. We note that given the low quality of the fit, we do not report the uncertainty on the best-fit parameters as they do not have any statistical significance. Instead, after applying the reconstruction, the best-fit becomes statistically accepted ($\mathrm{C-stat/dof} =1.02$), being consistent with the expected \cstat\ within $1.2\sigma$. The reconstructed spectra are shown in the top panel of Fig.~\ref{fig:groj1655}. The best-fit parameters are shown in Table~\ref{tab:fitgro}. The best-fit parameters deviate from the input values by less than 0.7\%. In the bottom panel of Fig.~\ref{fig:groj1655}, we show the confidence contours demonstrating the accuracy that could be obtained to determine the wind absorption parameters.

\begin{table}
\centering
\caption{Best-fit parameters obtained by fitting the simulated spectra of GRO~J155--40. The uncertainties correspond to an increase in \cstat\ by 1. We show the input parameters for comparison.}
\label{tab:fitgro}
\begin{tabular}{llll} 
\hline \hline
Parameter 	&	  Input 	&	Nominal	&	Reconstructed	 \\ \hline \\[-0.2cm]
$ N_{\rm H}~\rm (10^{23}~cm^{-2})$ 	&	 $4.0$ 	&	$4.04 $	&	 $3.994 \pm 0.002$	\\
$ \log \left( \xi/ {\rm erg~cm~s^{-1}} \right)$ 	&	 $4.5$ 	&	$4.5$	&	 $4.499 \pm 0.003$	\\
$ kT_{\rm in}$\,(keV) 	&	 $1.35$ 	&	$1.32 $	&	 $1.352 \pm 0.0003$	\\
$ \rm Norm_{\rm BB}$ 	&	 $483$ 	&	$79$	&	 $479.3 \pm 0.7$	\\
$ \Gamma$ 	&	 $3.5$ 	&	$3.23$	&	 $3.495 \pm 0.003$	\\
$ \rm Norm_{\rm pow}$ 	&	 $8.3$ 	&	$1.73 $	&	 $8.311 \pm 0.005$	\\ \hline \\[-0.2cm]
\cstat/dof	&		&	22143	&	1.02	\\
$\rm \Delta C~(\sigma)$	&		&	--	&	1.2	\\ \hline
\end{tabular}
\end{table}

\section{Summary and conclusions}
\label{sec:discussion}

In this paper, we presented a new technique that enables the analysis of defocused observations in which the PSF varies as a function of energy. We applied our method to the \athena/\xifu\ observations of bright sources, operated out of focus. We have tested the method for different flux levels, exposure times, and spectral shapes. We note that in this work we adopt a configuration (e.g., pixel size and event grading scheme) that differs from the one that will be adopted for the in-flight model. However, our results will not depend on the specific configuration as it is considered as an input to the method which will reconstruct the ARFs and the weights of each grade accordingly. We note that the method used to generate the weighted ARFs per event grade, presented in this work, be included in a future version of \sixte.

We have also demonstrated that the results of the reconstruction scheme, and thus the performance of the instrument depends heavily on the knowledge of the PSF and its energy dependency. In this work, we have used an analytical model of the PSF defined at 13 energies. We tested the uncertainties that are introduced by the calibration of the PSF as a function of energy, and the value of the PSF itself. These tests rely on a toy model to demonstrate the dependency of the method on the PSF knowledge. In the future, a physical/parametric model of the PSF will be available, which will allow us to perform end-to-end simulations in which we can fold-in all the mirror-instrument calibration uncertainties.

It is worth noting that the simulations presented in this work assume that the observed source is stable. However, it is well known that the bright sources that will be observed in the defocused mode (X-ray binaries for example) can be variable on very short timescales (on a millisecond timescale). The variability in flux and spectral shape will have to be accounted for by the method. Variability will alter the count rate detected by each of the \xifu\ pixels and the grade selection of the events consequently. Similarly, we do not address the effects of dithering that can affect the alignment of the PSF during the observation, that will need to be folded in the analysis. However, we note that for off-axis angles of up to $\sim 2\arcmin$ the defocused PSF will be identical to the on-axis version. In addition,  the positional accuracy of \athena\ will be of the order of $\sim 1\arcsec$, so this will not blur the off-axis PSF except, of course, the centre will be displaced by the off-axis angle and the area will drop slightly because of vignetting. However, all these effects will be taken into account during the analysis, and will be minor compared to the effect of event grading.

\begin{acknowledgements}
ESK acknowledges financial support from the Centre National d’Etudes Spatiales (CNES). JW and TD acknowledge partial funding from BMWi through Deutsches Zentrum f\"ur Luft- und Raumfahrt grant 50 QR 1903. 

This work makes use of AstroPy \citep{Astropy}, Matplotlib \citep{Matplotlib}, NumPy \citep{Numpy}, and SciPy \citep{Scipy}.
\end{acknowledgements}

\balance

\bibliographystyle{aa} 

%
%

\begin{appendix} 
\section{Plots}
\begin{figure*}
\centering
\includegraphics[width=0.45\linewidth]{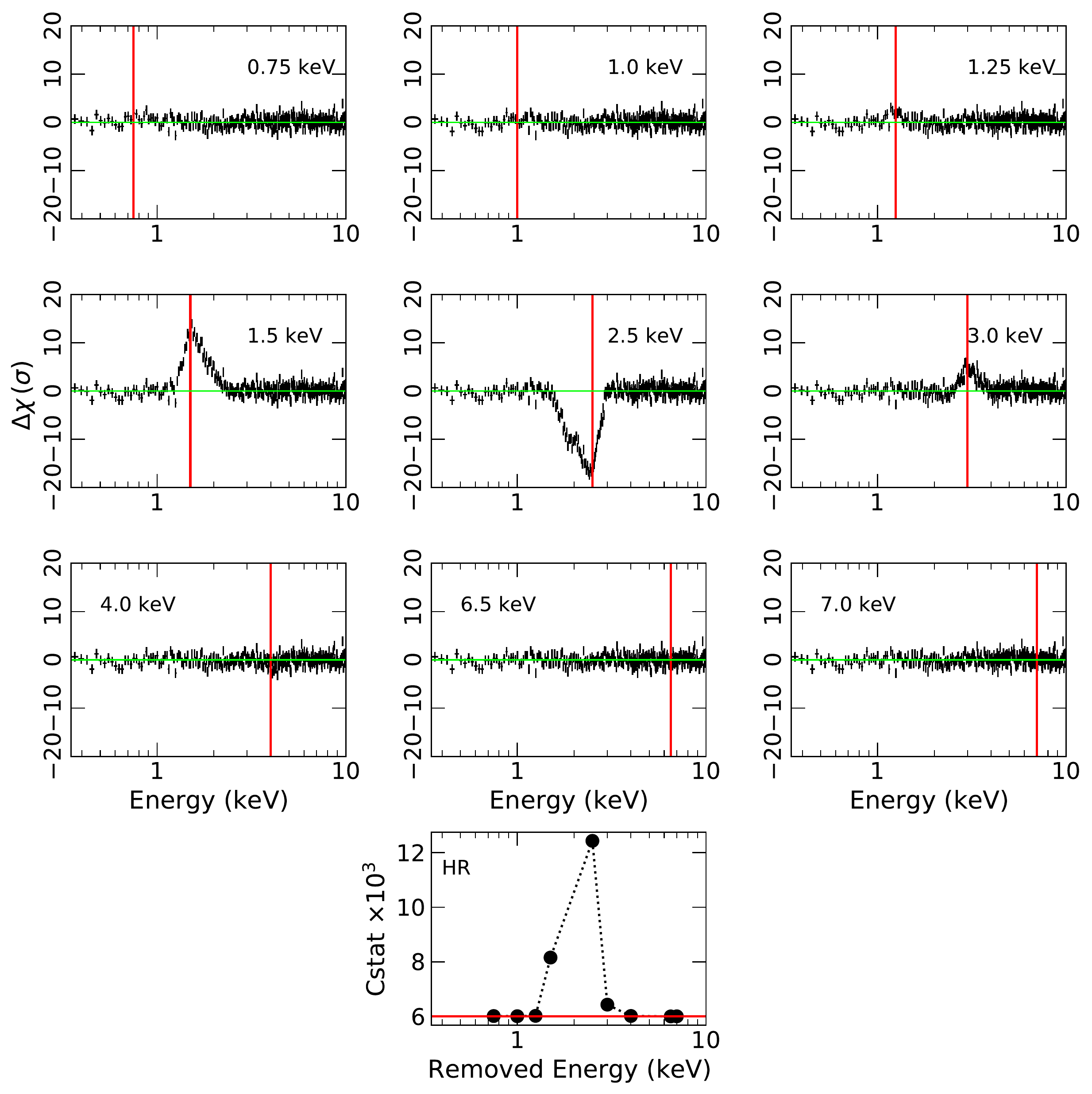}
\includegraphics[width=0.45\linewidth]{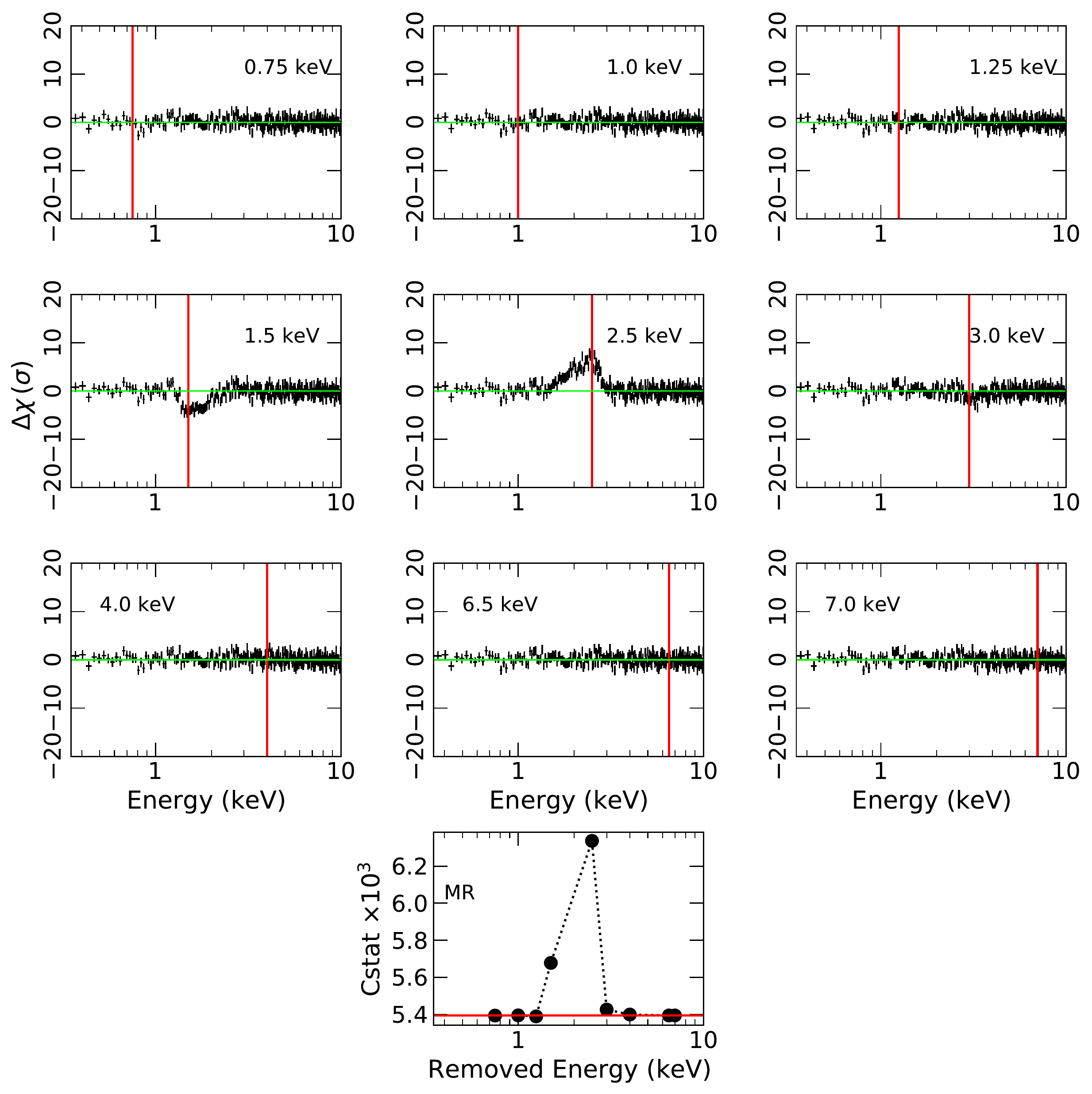}
\includegraphics[width=0.45\linewidth]{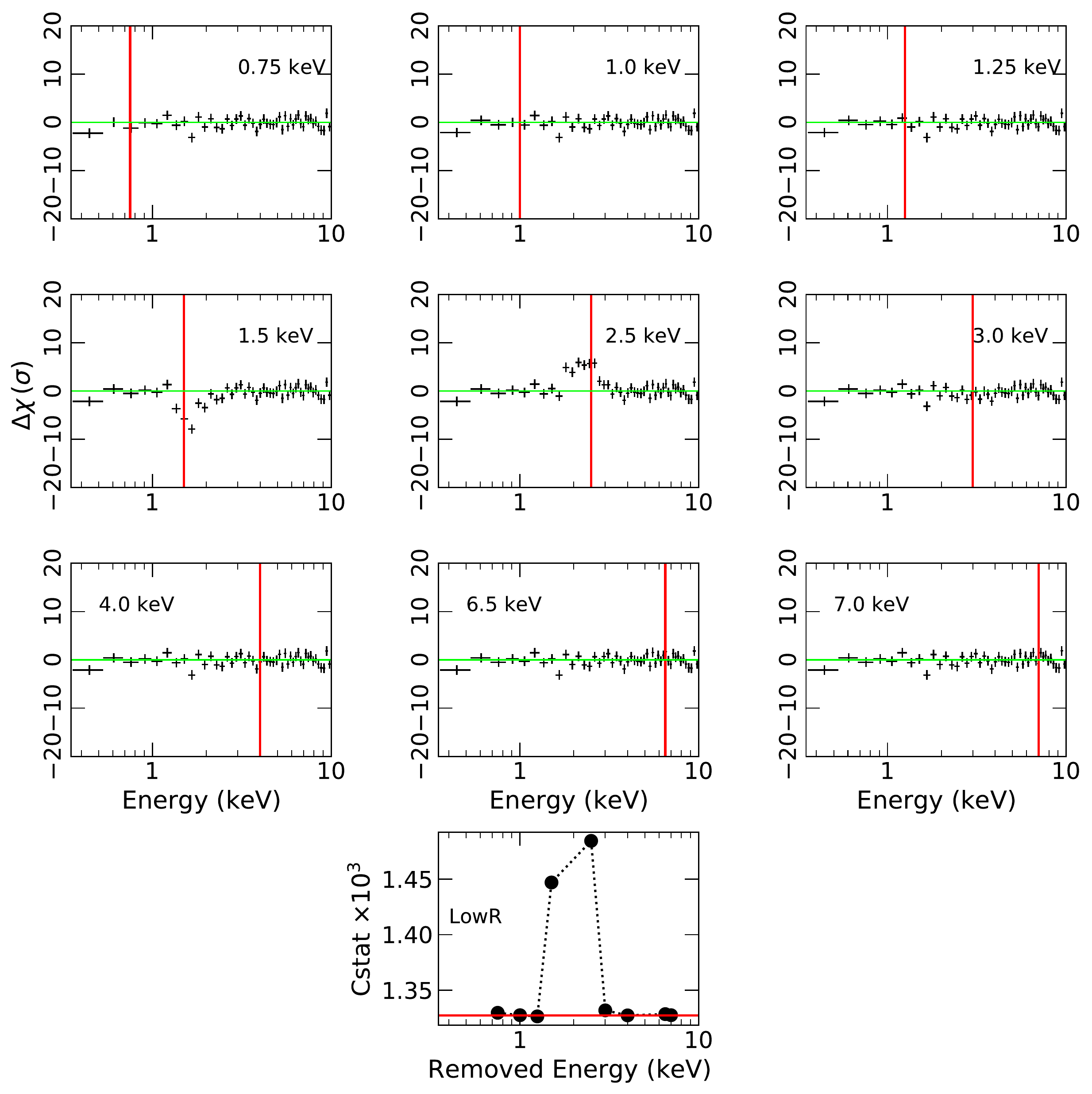}
\includegraphics[width=0.45\linewidth]{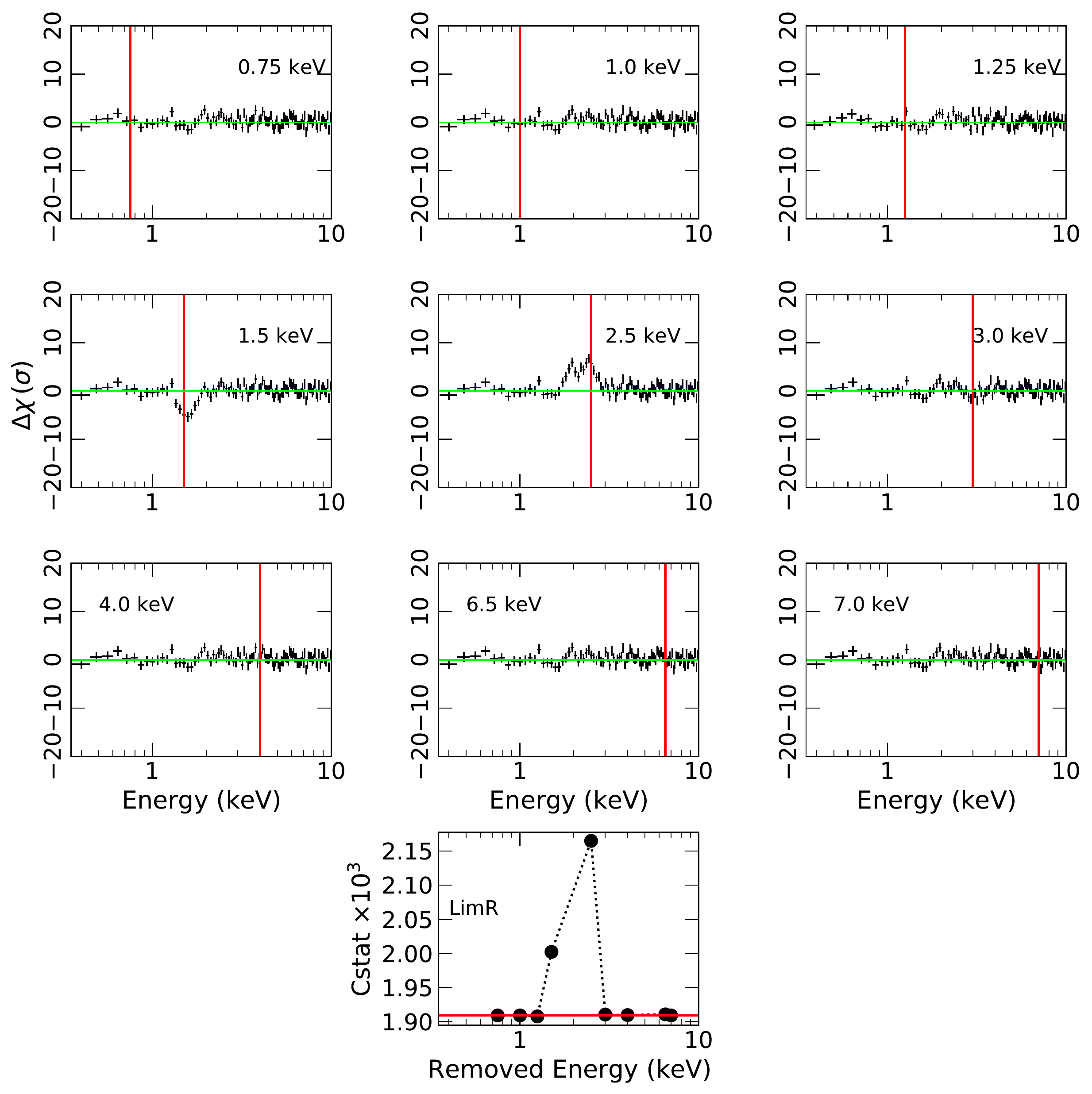}

\caption{Residuals for the different grades, assuming that the PSF is not known at a given energy (indicated by a vertical solid line) during the interpolation. The bottom panel shows how the \cstat\ changes as a function of the removed energy. The horizontal line in this panel corresponds to the \cstat\ value obtained by fitting the spectrum using all energies to reconstruct the ARF.}
\label{figapp:removing_energy}
\end{figure*}

\end{appendix}

\end{document}